\documentstyle[12pt]{article}
\newcommand{\cftnote}
{\renewcommand{\thefootnote}{\fnsymbol{footnote}}}
\newcommand{\resetftnote}{\setcounter{footnote}{0}}
\newcommand{\be}{\begin{equation}}
\newcommand{\ee}{\end{equation}}
\newcommand{\bea}{\begin{eqnarray}}
\newcommand{\eea}{\end{eqnarray}}
\def\bm#1{\mbox{\boldmath{$#1$}}}

\begin{document}

\begin{flushright}
UVA-94-09, hep-th/9403092\\
April 1995
\end{flushright}

\vspace{1mm}
{\Large \bf
\begin{center}
\cftnote
Unitarity of\\ The Realization of Conformal Symmetry\\ in The
Quantum Hall Effect%
\footnote{Research supported in part by CICYT (Spain) and FOM (The
Netherlands).}
\end{center}
}
\vspace{2mm}
\begin{center}
\cftnote
{\large Jos{\'e} Gaite%
\footnote{E-mail address: gaite@sisifo.imaff.csic.es}}
\\[4mm]
{\it Instituut voor Theoretische Fysica, University of Amsterdam,\\
Valckenierstraat 65, 1018 XE Amsterdam, The Netherlands}\\[2mm] and \\[2mm]
{\it Instituto de Matem{\'a}ticas y
F{\'\i}sica Fundamental, C.S.I.C.,\\ Serrano 123, 28006 Madrid,
Spain.}\\
\end{center}
\resetftnote

\vspace{2mm}

\begin{abstract}
We study the realization of conformal symmetry in the QHE as part of
the $W_\infty$ algebra. Conformal symmetry can be realized
already at the classical level and implies the complexification
of coordinate space. Its quantum version is not unitary. Nevertheless,
it can be rendered unitary by a suitable modification of its definition
which amounts to taking proper care of the quantum measure.
The consequences of unitarity for the Chern-Simons
theory of the QHE are also studied, showing the connection of
non-unitarity with anomalies.
Finally, we discuss the geometrical paradox of realizing
conformal transformations as area preserving diffeomorphisms.
\end{abstract}

\global\parskip 4pt

\section{Introduction}

The Quantum Hall Effect (QHE) is related to the existence of energy
levels with infinite (macroscopic) degeneracy for a system of
two-dimensional electrons in a perpendicular uniform magnetic field.
The infinite degeneracy immediately suggests the presence of an
infinite group of symmetry, which must have an important
role in the understanding of this effect.
Both the Integral and Fractional QHE are related to the incompressibility
of filled energy levels \cite{Laughlin}. Therefore, it was natural
to associate the symmetry group to the group of area preserving
diffeomorphisms, as was formally done in \cite{CapTZ,Iso}.
They showed that the symmetry of the classical system is indeed that
of area preserving diffeomorphisms, while at the quantum level
it becomes deformed to a well-known algebra, namely $W_\infty$.

On the other hand, alternative infinite symmetries,
namely, conformal or Kac-Moody, have also been shown to have a
role in the physics of the QHE. The 2d Coulomb gas, the simplest 2d CFT,
has been utilized since the original work of Laughlin \cite{Laughlin}.
Later, there were introduced vertex operator techniques \cite{Fubini}.
Independently, the study of edge waves led to a Kac-Moody symmetry
and, consequently, conformal symmetry \cite{Wen}.

While the importance of these symmetries is well established, it remains
to clearly understand their interrelations, in particular, that
between $W_\infty$ symmetry and conformal symmetry.
Cappelli {\em et al} observed that the $W_\infty$ algebra
contains a sub-algebra isomorphic to the algebra of
conformal transformations
and that this sub-algebra is indeed realized as conformal transformations
on analytic wave functions.
However, they point out that there are important differences with the
usual realization of this symmetry in 2d CFT \cite{BPZ}, for example,
the absence of negative modes.
Besides, they leave unexplained the appearance of conformal transformations
as a subgroup of area preserving diffeomorphisms, rather contradictory
from the geometrical point of view.

Our intention here is to clarify the nature of the realization
of conformal symmetry in the Hall effect. We consider in
section 2 the classical system, namely, one electron
(or several non-interacting electrons)
moving in a plane under a perpendicular uniform magnetic field.
For zero energy there is symmetry under area preserving
diffeomorphisms \cite{CapTZ}.
In the Hamiltonians formalism, the symmetry is realized as a group of
canonical transformations. If we admit complex canonical transformations,
the conformal transformations appear as a subgroup.
Upon quantization (section 3),
the algebra of canonical transformations becomes an
algebra of differential operators; the conformal subgroup corresponds
to those of first degree in derivatives. However, they are not self adjoint
and therefore do not generate unitary transformations in the Hilbert
space of holomorphic functions. The reason is that the transformation
of the integration measure has not been accounted for. We will show that
making the generator self adjoint (taking its real part) is equivalent to
adding the missing piece that takes care of the change of the measure under
conformal transformations. In section 4 are some pertinent
remarks on the peculiarities of the realization of conformal symmetry in
the QHE. Section 5 is dedicated to Chern-Simons theory. This theory
describes the gauge fluctuations in the Hall system and is
mathematically a functional version of the formalism introduced in
section 3. The realization of conformal symmetry is again related
through unitarity to the larger topological symmetry of the theory.
Furthermore, in this case a clear connection with anomalies arises.
The paper ends with a discussion on the relation between conformal
transformations and area-preserving diffeomorphisms in the QHE.

\section{Symmetry of the classical system.}

Let us briefly recall the constrained Hamiltonians formulation of one electron
with zero energy in a plane under a perpendicular uniform magnetic field
\cite{Dunne,CapTZ}. The Hamiltonians
\be
H = {1 \over 2}\, m\, {\bm v}^2
\ee
is constrained to be zero by imposing that the velocity ${\bm v}$,
\be
v^a ={1 \over m} \left(p_a + {B\over 2}\, \epsilon_{ab}\,x^b\right),
\ee
vanish. One usually defines complex phase space variables,
\be
a = {1 \over 2} \left(-v_2 + i\,v_1 \right)
\ee
and $b$, which results from $a$ by reversing the sign of the magnetic
field $B$. The constraints are then expressed as
\be
a = a^* = 0,
\ee
which imply that
\be
p_y = x,~~~p_x = -y.
\ee
The four-dimensional phase space effectively reduces to a two-dimensional
phase space. This fact is the first sign of the dynamics
being reduced to the edge of the sample.
In the reduced phase space, the value
of $b$ only depends on the coordinates $x$ and $y$,%
\footnote{To simplify, we will
use units such that $c= \hbar = 1$, $m=1$ and $B=2$, implying that
the magnetic length is also one.}
\be
b = x - i\,y \equiv z^*,~~~b^* = x + i\,y \equiv z.
\ee
An arbitrary function of $x$ and $y$ generates canonical transformations
that leave the (null) Hamiltonians unchanged and are therefore symmetries.
Furthermore, they preserve the element of area $dx \wedge dy$,
which is the symplectic form, and
constitute the algebra of area preserving diffeomorphisms,
called $w_\infty$.

One can consider as generators arbitrary functions $F\left(z, z^*\right)$,
that is not necessarily real. This implies complexifying $x$ and $y$.
Hence, according to the usual convention we use
$\bar{z}$ instead of $z^*$.\footnote{Note that $\bar{z}= x - i\,y$ whereas
$z^* = x^*- i\,y^*$.}
The transformations of $z$ and $\bar{z}$ are still given by
\bea
\delta_F\,z = i\,{\partial F \over \partial \bar{z}}  \label{transfz} \\
\delta_F\,\bar{z} = -i\,{\partial F \over \partial z} \label{transfbarz}
\eea
but now $z$ and $\bar{z}$ transform independently. An interesting example
is given by
\be
F\left(z, \bar{z}\right) = -i\,\bar{z}\,\epsilon(z),   \label{F_holo}
\ee
which yields
\bea
\delta_F\,z = \epsilon(z),  \\
\delta_F\,\bar{z} = - \partial_z \epsilon(z)\, \bar{z}.
\eea
Recalling that $\bar{z}$ plays the role of the conjugate momentum of $z$,
since their Poisson bracket is
\be
\left\{z, \bar{z}\right\}_{P.B.} = i,   \label{fundPB}
\ee
this transformation is the complex analogue of a point transformation:
$z$ undergoes an arbitrary (holomorphic) diffemorphism and $\bar{z}$
transforms as a co-vector.

Given the dual role of $z$ and $\bar{z}$, we can also consider
\bea
F(z, \bar{z}) = -i\,z\,\bar{\epsilon}(\bar{z}), \\
\delta_F\,z = - \partial_{\bar{z}}\bar{\epsilon}(\bar{z})\,z, \\
\delta_F\,\bar{z} = \bar{\epsilon}(\bar{z}).
\eea
Thus, we have holomorphic and anti-holomorphic transformations as subgroups
of the complex area preserving diffeomorphisms.
However, holomorphic transformations of $z$ affect $\bar{z}$ as well and
vice-versa anti-holomorphic transformations,
thence they do not commute, unlike those standard in 2dCFT.

\section{Realization of conformal symmetry in the quantum system}

In the quantum system one has operators, $a, a^{\dag}, b$ and
$b^{\dag}$, in place of phase space variables.
The Hamiltonians
\be
H = 2\,a^{\dag}a + 1
\ee
commutes with any power of $b$ and $b^{\dag}$.
The constraints are implemented a la Gupta-Bleuler,
\be
a\,\Psi = 0,
\ee
meaning that the state $\Psi$ belongs to the lowest Landau level. This
condition implies that the wave function adopt a particular form,
\be
\Psi(x, y) = \psi(z)\, {\rm e}^{-{1 \over 2}\,z\bar{z}},
\ee
which is called the holomorphic or coherent state representation. The
operators $b$ and $b^{\dag}$ act on holomorphic wave functions as\footnote
{This representation can alternatively be obtained
by quantizing the constrained
system with Poisson bracket (\ref{fundPB}).}
\bea
b\,\psi(z) = \partial_z\psi(z),  \\
b^{\dag}\,\psi(z) = z\,\psi(z).
\eea
Hence, the infinite symmetry is generated by
arbitrary differential operators in $z$.
A convenient basis is given by
\be
{\cal L}_{n,m} = z^{n+1}\,\partial_z^{m+1},~~~~n, m \geq -1.   \label{L}
\ee
They satisfy the commutation relations of the celebrated
$W_\infty$ algebra, a deformation of $w_\infty$.

A notable sub-algebra occurs when $m=0$, namely,
the algebra of holomorphic transformations
\be
\delta\psi(z) = z^{n+1}\,\partial_z\psi(z).  \label{dhwf}
\ee
This algebra can be regarded as the quantum representation
of (\ref{F_holo}). We remarked there that the function $F$ in
(\ref{F_holo}) is complex and does
not generate transformations in the real $(x,y)$ plane.
Its quantum version, a linear combination of operators ${\cal L}_{n,0}$,
is not self adjoint, as one
can easily see. Let us expound this point.
We know after Dirac that canonical transformations are
represented in Quantum Theory as unitary transformations of
the Hilbert space.\footnote{True at least locally.}
The generator of a canonical transformation
becomes a self-adjoint operator.
In particular, $w_\infty$ comprises all the canonical transformations
of the reduced phase space; its quantum counterpart, $W_\infty$,
is the algebra of unitary transformations on the lowest Landau level
\cite{Iso}. However, the complexification necessary to include
holomorphic transformations spoils unitarity and
the operators ${\cal L}_{n,0}$ do not provide
a unitary representation of conformal symmetry.

To look further into this problem, we shall first consider the unitarity of
canonical transformations and, in particular, point transformations in
a more general setting; namely, an arbitrary quantum system, which we
take one dimensional for simplicity. There is one coordinate $x$ and one
momentum $p$; states are given by their wave functions $\psi(x)$.
Point transformations are generated by
\be
F(x, p) = f(x)\,p,
\ee
with
\bea
\delta_F\,x = f(x),  \\
\delta_F\,p = -f'(x)\,p.
\eea
Since in the coordinate representation $p = -i\,\partial_x$, one may think
that
\be
\delta_F\,\psi(x) \equiv i\,F\,\psi(x) = f(x)\,\partial_x\psi(x).
\label{dwf}
\ee
Hence, $\psi(x)$ transforms as a scalar. To see the change of the norm
\be
\langle \psi \mid \psi \rangle = \int {\rm d}x\:\psi^*(x)\,\psi(x),
\label{norm}
\ee
we calculate
\be
\delta_F \left(\psi^*(x)\, \psi(x)\right) =
f(x)\,\partial_x(\psi^*(x)\, \psi(x))
\ee
and find a non-null value.

It is natural because we have not taken
into account the change of the element of volume in (\ref{norm}).
On the other hand, as the alert reader has probably noticed,
the operator $F$ in (\ref{dwf}) is not self adjoint. We must consider
instead
\be
\hat{F} = {1 \over 2}\left(F + F^{\dag} \right) =
{1 \over 2}\left(f(x)\,(-i\,\partial_x) + (-i\,\partial_x)\,f(x)\right),
\label{hatF}
\ee
and
\be
\delta_{\hat{F}}\, \psi(x) = f(x)\,\partial_x\psi(x) +
{1 \over 2}\,\partial_xf(x)\,\psi(x).    \label{ndwf}
\ee
Now,
\be
\delta_{\hat{F}} \left(\psi^*(x)\, \psi(x)\right) =
f(x)\,\partial_x\left(\psi^*(x)\, \psi(x)\right) +
\partial_xf(x)\left(\psi^*(x)\, \psi(x)\right),  \label{dswf}
\ee
which accounts for the change of the element of volume, so that
\be
\delta_{\hat{F}}\,\langle \psi \mid \psi \rangle = 0.
\ee
This extends to any canonical transformation: One cannot just take
the action of $F$ on wave functions. One must also take care
of the element of volume and this is automatically done by taking
the self-adjoint part of $F$.

Let us return to transformations of holomorphic wave functions.
Comparing (\ref{dhwf}) with (\ref{dwf}), one may think by analogy that what
is missing in the former to be self adjoint is the change of the
corresponding element of volume. Let us recall the form of the norm in
the holomorphic representation,
\be
\langle \psi \mid \psi \rangle =
\int {{\rm d}z\,{\rm d}{\bar z}\over 2\pi i}\,
{\rm e}^{-z{\bar z}}\:{\bar\psi}({\bar z})\,\psi(z). \label{hnorm}
\ee
The element of volume or holomorphic measure is
\be
{\rm d}\mu = {{\rm d}z\,{\rm d}{\bar z}\over 2\pi i}\,{\rm e}^{-z{\bar z}}.
\label{measure}
\ee

Every operator has a holomorphic-antiholomorphic kernel \cite{I-Z}.
A classical function and the kernel of its corresponding quantum
operator coincide when the anti-normal ordering prescription
\cite{G-J} is assumed. In particular, the kernel of a holomorphic
transformation of wave functions, (linear combination of (\ref{dhwf}),)
is precisely (\ref{F_holo}). Since this operator is not self adjoint,
we define\footnote{$\bar{\epsilon}$ is the function complex conjugate
to $\epsilon$.}
\be
\hat{F} = F + {\bar F} = -i\left(\bar{z}\,\epsilon(z) -
z\,\bar{\epsilon}\left(\bar{z}\right) \right).
\label{hatFz}
\ee
When $x$ and $y$ are real, $\hat{F}$ is a real function and is
as well the kernel of a self-adjoint operator.
The coordinate transformation produced by this function according to
(\ref{transfz}, \ref{transfbarz}) is
\bea
\delta_{\hat{F}}\, z = \epsilon(z) -
z\,\partial_{\bar{z}} \bar{\epsilon}(\bar{z}), \label{ntz}\\
\delta_{\hat{F}}\, {\bar{z}} = \bar{\epsilon}(\bar{z}) -
\bar{z}\,\partial_z \epsilon(z),  \label{ntzb}
\eea
which is a perfect area-preserving diffeomorphism of real $x$ and $y$.
The quantum transformation is given by
\be
\delta_{\hat{F}}\,\psi(z) = \delta_F\,\psi(z) + \delta_{\bar F}\,\psi(z),
\ee
with
\be
\delta_F\,\psi(z) \equiv i\,F\,\psi(z) = \bar{z}\,\epsilon(z)\,\psi(z) =
\partial_z\left(\epsilon(z)\,\psi(z) \right) =
\partial_z\epsilon(z)\,\psi(z) + \epsilon(z)\,\partial_z\psi(z).
\ee
Similarly,
\be
\delta_{\bar F}\,{\bar\psi}(\bar{z}) =
\partial_{\bar{z}}\bar{\epsilon}(\bar{z})\,{\bar\psi}({\bar z}) +
\bar{\epsilon}(\bar{z})\,\partial_{\bar{z}}{\bar\psi}({\bar z}).
\ee
Therefore, we can write
\bea
\label{ndhwf}
\delta_{\hat{F}}\,\psi(z) =
(\partial_z\epsilon(z) + \epsilon(z)\,\partial_z
- z\,\bar{\epsilon}(\bar{z})) \,\psi(z),         \\
\label{ndawf}
\delta_{\hat{F}}\,{\bar\psi}(\bar{z}) =
(\partial_{\bar{z}}\bar{\epsilon}(\bar{z}) +
\bar{\epsilon}(\bar{z})\,\partial_{\bar{z}} -
\bar{z}\,\epsilon(z))\,{\bar\psi}(\bar{z}),
\eea
and, consequently,
\bea
\delta_{\hat{F}}\left({\bar\psi}(\bar{z})\,\psi(z)\right) =
\left(\epsilon(z)\,\partial_z +
\bar{\epsilon}(\bar{z})\,\partial_{\bar{z}}\right)
\left({\bar\psi}(\bar{z})\,\psi(z)\right) +        \nonumber \\
\left(\partial_z\epsilon(z) - z\,\bar{\epsilon}(\bar{z}) +
\partial_{\bar{z}}\bar{\epsilon}(\bar{z}) - \bar{z}\,\epsilon(z)\right)
\left({\bar\psi}(\bar{z})\,\psi(z)\right).     \label{dshwf}
\eea
The latter equation has the form of (\ref{dswf}). The first term expresses
the variation under the conformal transformation generated
by $\epsilon(z)$ and we can expect that the
second term accounts for the change of the measure (\ref{measure}).
Let us compute it,
\bea
\delta_{\epsilon}\,{\rm d}\mu =
\delta_{\epsilon}({\rm d}z\,{\rm d}\bar{z})\,{\rm e}^{-z{\bar z}}
+ {\rm d}z\,{\rm d}\bar{z}\,\delta_{\epsilon}{\rm e}^{-z{\bar z}} =
\nonumber\\
(\partial_z\epsilon(z)\,{\rm d}z\,{\rm d}\bar{z} +
{\rm d}z\,\partial_{\bar{z}}\bar{\epsilon}(\bar{z})\,{\rm d}\bar{z})\,
{\rm e}^{-z{\bar z}}
+ {\rm d}z\,{\rm d}\bar{z}\,(-z\,\bar{\epsilon}(\bar{z})
- \bar{z}\,\epsilon(z))\,
{\rm e}^{-z{\bar z}} =          \nonumber   \\
\left(\partial_z\epsilon(z) - z\,\bar{\epsilon}(\bar{z}) +
\partial_{\bar{z}}\bar{\epsilon}(\bar{z}) - \bar{z}\,\epsilon(z)\right)
{\rm d}\mu.
\label{dm}
\eea
Hence we have full confirmation.

In the real case, adding $F^{\dag}$ (\ref{hatF}), is equivalent
to give a definite (symmetrical) order to $x$ and $p$. It introduces
in (\ref{ndwf}) the derivative of the diffeomorphism generator $f(x)$,
which expresses the change of d$x$. In the present holomorphic case
there are two extra pieces in $\delta_{\hat{F}}\,\psi(z)$ (\ref{ndhwf}):
The anti-normal ordering produces $\partial_z\epsilon(z)$ , expressing
the change of d$z$, whereas ${\bar F}$ produces the terms that
accounts for the change of the non-holomorphic factor
${\rm e}^{-z{\bar z}}$
in the measure (\ref{measure}). An identical statement can be made
for the anti-holomorphic component (\ref{ndawf}).

One can perceive here an analogy with the modern philosophy of anomalies
or, more generally, with the realization of non-linear transformations
in quantum theory. It is not sufficient to translate the classical realization
of a symmetry to quantum states or operators; the measure in the path integral
must also be taken into account. Failure to do this
appears as lack of unitarity. The case of the non-linear $\sigma$-model is
paradigmatic.\footnote{See the introduction of \cite{Hata}
for a brief review of the $\sigma$-model in this context.}
The conformal symmetry of the Classical HE, which is
translated as the quantum operator $F$, must be supplemented by terms that
take care of the measure, contained in ${\bar F}$, curing at the same time
the unitarity problem.

\section{A closer look at conformal symmetry}

So far, the function $\bar{\epsilon}(\bar{z})$ has been taken to be
the complex conjugate of $\epsilon(z)$. However, we can consider
in (\ref{hnorm}) $z$ and $\bar{z}$ as independent complex integration
variables. Then it is clear that one can perform on them independent
arbitrary holomorphic transformations.\footnote{This holds at least
for transformations close to the identity, which do not spoil the
convergence of the integral.} Hence, we can enlarge our symmetry
to the entire group of holomorphic or anti-holomorphic transformations,
as in standard 2d CFT. This might seem to contradict our previous
assertion that the classical action of holomorphic and
anti-holomorphic transformations do not commute with each other.
The cause of this non-commutativity was that holomorphic, say,
transformations
($\bar{\epsilon}(\bar{z}) = 0$) affect $\bar{z}$ as well.
In contrast, although holomorphic transformations act on
${\bar\psi}(\bar{z})$ as
\be
\delta_F\,{\bar\psi}(\bar{z}) = -\bar{z}\,\epsilon(z)\,{\bar\psi}(\bar{z}),
\ee
it is just what corresponds to the factor ${\rm e}^{-z{\bar z}}$
in the measure; the function itself does not change.

It seems that we have essentially the same realization that
in standard 2d CFT. $\psi(z)$ (respectively, ${\bar\psi}(\bar{z})$)
transforms as a scalar function under holomorphic (resp., anti-holomorphic)
transformations. However, the unitarity conditions are different:
Standard 2d CFT is constructed in 2d Minkowski space; hence,
$z = x - t$ and $\bar{z} = x + t$ are assumed real and so are
the generators $\epsilon(z)$ and $\bar{\epsilon}(\bar{z})$.
Upon Fourier mode expansion,\footnote{Valid for periodic $x$, that is,
the topology of a cylinder.}
\bea
\epsilon(z) = \sum_{-\infty}^{\infty} \epsilon_n\,{\rm e}^{i\,n\,z},~~~
l_{n} = {\rm e}^{i\,n\,z}\,\partial_z,         \label{Fme} \\
\bar{\epsilon}(\bar{z}) =
\sum_{-\infty}^{\infty} \bar{\epsilon}_n\,{\rm e}^{i\,n\,\bar{z}},~~~
\bar{l}_n = {\rm e}^{i\,n\,\bar{z}}\,\partial_{\bar{z}},
\nonumber
\eea
one obtains the usual conditions,
\be
l_n^{\dag} = l_{-n},~~~%
\bar{l}_n^{\dag} = \bar{l}_{-n}.
\ee
They are commonly expressed in another coordinate system,\footnote
{Corresponding to mapping the cylinder onto the punctured plane.}
$w = {\rm e}^{i\,z}$: Fourier mode expansions become Laurent expansions
and $w^* = w^{-1}$.

In the QHE, we are in a genuinely euclidean 2d space. Hence, $\bar{z}$
is the complex conjugate of $z$ and $\bar{\epsilon}(\bar{z})$
the complex conjugate of $\epsilon(z)$. Therefore, we directly use
a Taylor expansion and unitarity requires
\be
l_n^{\dag} = \bar{l}_n
\ee
instead. This unitarity condition mixes the holomorphic and
anti-holomorphic sectors and is therefore unrelated to the
inner product of standard 2d CFT, where those sectors are independent.
However, that mixing was to be expected for the inner product
provided by the norm in the holomorphic representation (\ref{hnorm}).

As regards to the comparison with standard 2d CFT, there is another
point that also deserves attention:
The absence of negative modes, according to~(\ref{L}) \cite{CapTZ}.
In contrast to their conclusion, our opinion is that that absence
makes no essential difference.
The argument relies on the previous paragraph.
Although in standard 2d CFT the generator of transformations
is allowed to be singular at the origin, this is a coordinate dependent
statement: It is singular in $w$ but not in $z$. We can further Taylor
expand in~(\ref{Fme}),
\be
l_{n} = {\rm e}^{i\,n\,z}\,\partial_z =
\sum_{k=0}^{\infty}{1 \over k!}\,(i\,n\,z)^k\,\partial_z,
\ee
showing that the positive modes suffice in this coordinate
system.\footnote{We recall that the type of expansion is
motivated by global considerations and should not affect local objects.}

\section{Gauge fluctuations and Chern-Simons theory}

To treat fluctuations of the gauge field we shall now include it as
a dynamical field in the system of many interacting electrons. Then
it fluctuates around the value of the external magnetic field $B$.
The Chern-Simons interaction is known to be dominant at long distance
in electrodynamics
in $2+1$ dimensions \cite{Fro}. Its topological character implies that
the dynamics is confined to the boundary of the manifold (usually
a cylinder). Therefore it directly relates to the edge-current picture
of the QHE. It is also noteworthy that these theories are equivalent
to theories of chiral bosons living at the edge of the sample \cite{Bala2},
thus providing a concrete Lagrangian that realizes 2d conformal symmetry.
{}From a more mathematical standpoint, non-abelian Chern-Simons theory
was formulated by Witten as a covariant 3d theory of knots. In his paper
\cite{Witten} the connection with 2d CFT is already made. This connection
was explained in subsequent papers by other authors.
In particular, the application to the QHE, based on the group $U(1)$,
was studied in \cite{Bala1}.
Here I intend to sketch how conformal symmetry arises in Chern-Simons theory
and to show that the issue of unitarity in Chern-Simons theory is related
not to unitarity in 2d CFT but to modular invariance. Besides, it is also
worthwile to point out the relation between unitarity and the anomaly
of Chern-Simons theory in the presence of a boundary.

The observables of Chern-Simons theory are given by functional integration
of the appropriate classical functions weighted with the exponential
of the action. The connection with 2d CFT is made through canonical
quantization (or operator formalism) and elimination of
the gauge degrees of freedom \cite{Witten}. It is carried through
by first cutting the 3d manifold $M$ along a Riemann surface $\Sigma$ to
establish the classical phase space, which produces the Hilbert space.
The operation is best made in the holomorphic representation
for there are constraints (the Gauss law).
In this representation the partition function adopts
the form of a scalar product
which is a functional version of (\ref{hnorm}) \cite{BN,Laba}
\be
Z = \langle \Psi \mid \Psi \rangle =
\int {{\rm D}A_z\,{\rm D}{A_{\bar z}}\over 2\pi i}\,
\exp\left(-\int {\rm d}^2z\,A_z\,{A_{\bar z}}\right)\:
{\overline\Psi}\left[{A_{\bar z}}\right]\,\Psi\left[A_z\right].
\label{fhnorm}
\ee
This functional integral is over configurations on $\Sigma$.
The wave functionals can be obtained by functional integration
of the Chern-Simons action with a boundary term \cite{Laba},
\be
\Psi\left[A_z\right] = \int {\rm D}A\:
\exp\left(i\int_M A \wedge {\rm d}A +
{1\over 2}\int_\Sigma {\rm d}^2z\,A_z\,{A_{\bar z}}\right). \label{fhwf}
\ee
After removing gauge equivalent configurations in (\ref{fhnorm})
one is left with
an integral over the moduli space of $U(1)$ flat connections.
On general grounds, this integral represents the corresponding
WZW model and has conformal symmetry \cite{Witten,BN,Laba}.

Concretely, one can express the partition function with $n$
Wilson lines piercing $\Sigma$ at points $z_1, \dots, z_n$
as a correlator of vertex operators
\bea
Z(L_1, \ldots, L_n) =
\langle V_{\alpha_1}(z_1, {\bar z}_1) \cdots
V_{\alpha_n}(z_n, {\bar z}_n)\rangle_{A_{\rm cl}} =   \\
\sum_{K\bar L} h_{K{\bar L}}\,
{\cal F}_K(z_1, \dots, z_n, \tau; A_z^{\rm cl})\,
{\overline{\cal F}}_{\bar L}(\bar z_1, \ldots, {\bar z}_n, {\bar\tau};
A_{\bar z}^{\rm cl});      \label{ZWl}
\eea
$z_i$ are (moduli) variables for the punctures and
$\tau$ label other possible moduli of the Riemann surface.
The right-hand side is the conformal-block decomposition of
the correlator.%
\footnote{For a review on 2d CFT see, for example, \cite{spa}.}
We mean these Wilson lines to represent physical particles, whereby
the background classical field $A_{\rm cl}$ is produced by these particles.
The matrix $h_{K\bar L}$
encodes the data that determine a particular 2d CFT.

As is well known, the partition function $Z(L_1, \ldots, L_n)$
can be exactly calculated in the abelian case
since the functional integral is gaussian, to give
\be
Z(L_1, \ldots, L_n) =
\exp \left( -{i \over 2}\,\sum_{k=1}^n\int {\rm d}x_k\,A_k^{\rm cl}\right),
\ee
which for closed paths yields the linking number (once normalized).
To quantize the physical particles we consider the path integral over
trajectories $x(t)$
(with only one particle for simplicity)
\be
Z' = \int {\rm D}x\:
\exp\left( -{i \over 2}\,\int {\rm d}x\,A^{\rm cl}\right).
\ee
In the QHE $A_i^{\rm cl}$ is to be taken as a mean field, which must
coincide with the external field,
\be
\langle B \rangle =  B.
\ee

Thus one arrives to the Lagrangian formulation of the QHE, alternative
to the Hamiltonian formulation exposed in section 2. The path integral
admits a holomorphic representation, analogous to (\ref{fhnorm}):
\be
Z' = \int_{\Sigma} {{\rm d}z\,{\rm d}{\bar z}\over 2\pi i}\,
{\rm e}^{-z{\bar z}}\:{\bar\psi}({\bar z})\,\psi(z), \label{Zp}
\ee
with
\be
{\psi}(z) = \int {\rm D}x\: \exp\left(i
\int {\bm x} \wedge {\rm d}{\bm x}
+ {1\over 2}\,{z{\bar z}} \right),  \label{piwf}
\ee
where the action
\be
S = \int {\bm x} \wedge {\rm d}{\bm x} =
\int {\rm d}t\,(x_1\,\dot{x}_2-x_2\,\dot{x}_1)
\ee
is the magnetic flux or area enclosed by the trajectory
projected on $\Sigma$. Note that it is a topological invariant which plays
a similar r{\^o}le to the Chern-Simons action \cite{Dunne,CapTZ}.
The identification of ${1\over 2}\,{z{\bar z}}$ as a boundary term is
readily made. When performing the path integral (\ref{piwf}) no boundary
condition is imposed on ${\bar z}$. Therefore, the variation of the
action has a boundary component:
\bea
\delta S =
2\int \delta{\bm x} \wedge {\rm d}{\bm x} -
({\bm x} \wedge \delta{\bm x})_{t=0},\\
{\bm x}(0) \wedge \delta{\bm x}(0) \equiv {1\over {2\,i}}\,(z\,\delta{\bar z}
- {\bar z}\, \delta z) = {1\over {2\,i}}\, z\,\delta{\bar z}.  \label{bcomp}
\eea
This component is canceled by the variation of the boundary term.
Incidentally, the same term cancels the boundary component in
${\bar\psi}(\bar{z})$, but it involves two extra minus signs.

In the many-particle case we have in the integrand of
(\ref{Zp}) Laughlin wave functions. Chern-Simons mean field
equations impose that the average particle density is
proportional to the external magnetic field,
\be
\langle \rho \rangle =  B,         \label{mft}
\ee
hence constant.
Since $Z'$ is just the path
integral over Wilson lines of the partition function (\ref{ZWl}),
we see that the conformal blocks ${\cal F}_K$ are to be identified
with the holomorphic part of Laughlin's wave functions \cite{MoRe}.

In principle, we would like the correlators of Chern-Simons theory
to be independent of the complex structure chosen for the Riemann surface,
given the topological character of that theory.
We should therefore demand that the partition function $Z$ be independent
of the moduli parameters ($z_i$, $\tau$).
This turns out to be impossible to achieve
in the quantum theory because of regularization problems. We have to deal
with an anomaly. The theory is no more independent
of the metric on $\Sigma$ but depends on the conformal factor.
This regularization problem also appears as the necessity of
``framing" the Wilson lines, introducing a dependence on local coordinates
that amounts to the same thing. This is the way by which conformal
invariance comes in this picture of the QHE. The topological nature is
now manifested as the invariance under those transformations of the
moduli parameters that do not alter the Riemann surface. In other words,
the partition function (or the correlators) must be modular invariant.
This is the crucial condition on
the matrix $h_{K\bar L}$ (modular invariance) and is sufficient
to determine it. Given that the partition function is the holomorphic norm
(\ref{fhnorm}),
we can appreciate that modular invariance is a consequence of
the Chern-Simons unitarity condition.

Returning to the relation of non-unitarity and anomalies already remarked upon
at the end of section 3, we can see it more clearly in Chern-Simons theory
since it has actual gauge invariance.
The gauge transformation of wave functionals is given by
\be
\delta_\chi\Psi\left[A_z\right] =
\int_{\Sigma}\delta_\chi A_z\,A_{\bar z}\,\Psi =
\int_{\Sigma}{\delta\Psi \over \delta A_z} \,\delta_\chi A_z =
\int_{\Sigma}{\delta\Psi \over \delta A_z} \,\partial_z\chi     \label{fdwf}
\ee
and is anomalous, namely, it is such that $\delta_\chi Z \neq 0$.
The reason is again the presence of a non-trivial integration
measure in $Z$ (\ref{fhnorm}). A general non-linear transformation
of the wave functional, with $\delta A_z$ arbitrary in (\ref{fdwf}),
is non unitary. Similarly to (\ref{ndhwf}), we must add two terms
in (\ref{fdwf}),
\be
\delta\Psi\left[A_z\right] =
\int_{\Sigma}\left(\delta A_z\,{\delta \over \delta A_z} +
{\delta \over \delta A_z}(\delta A_z) -
A_z \,\delta A_{\bar z}\right)\Psi\left[A_z\right].   \label{cfdwf}
\ee
The term with the derivative of $\delta A_z$ is absent in
the particular case of a gauge transformation.
The other extra term is the variation of the exponential
part of the measure. As already said, this exponential arises
as a boundary term in the functional integral (\ref{fhwf})
for the wave functional to cancel the boundary component in the
equation
(compare with (\ref{bcomp}))
\be
\delta \int_M A \wedge {\rm d}A =
2\int_M \delta A \wedge {\rm d}A -
\int_\Sigma A \wedge \delta A.
\ee
When the variation is due to a gauge transformation $\delta_\chi A_z =
\partial_z\chi$ that extra term can be interpreted as the ``gauge anomaly''
of Chern-Simons in a manifold with boundary \cite{Bala2},
It is possible to see that it is precisely the one
required to cancel the chiral anomaly in 2 dimensions \cite{Bala2,FroStu}.

One can observe that the topological anomaly of Chern-Simons theory
can be attributed to the same boundary term (or to the quantum
measure). The partition function
as a whole must be invariant under diffeomorphisms and it indeed yields
topological invariants, like the linking number. However, the
wave functions (conformal blocks) are only conformal invariant.
It is the boundary term what introduces in the wave function (\ref{fhwf})
a dependence on the conformal factor of the 2d metric. If we account for it
as in (\ref{cfdwf}) we recover topological invariance.

\section{Discussion}

We have seen that a sub-algebra of $W_{\infty}$ can be realized as conformal
transformations, despite being also realized as area preserving
diffeomorphisms in the classical system. This dual role may seem puzzling,
given the very different geometrical nature of these two
types of transformations, but
there is no contradiction. One can further confirm that the former
realization becomes
the latter in the classical limit, when
$$
\langle \epsilon(z) \rangle = \epsilon(\langle z \rangle).
$$
However, this confirmation is not very enlightening.
It may be better to resort to
particular examples of $\epsilon(z)$ to gain some insight.

The conformal
transformation that most prominently modifies the area is a dilation,
$\epsilon(z)= \epsilon\,z$ with $\epsilon$ real. Interestingly,
$\hat{F}$ vanishes for it (\ref{hatFz}),
as well as the corresponding classical diffeomorphisms
(\ref{ntz}, \ref{ntzb}).
Nevertheless, $\psi(z)$ transforms as it should.
The best way to understand it is within Laughlin's original philosophy
\cite{Laughlin}, in which the non-holomorphic factor in the wave functions
is interpreted as a background potential for the 2d Coulomb gas.
In some sense, the dilation of the holomorphic wave function
is compensated for by
a similar dilation of the neutralizing background charge, represented by
the factor ${\rm e}^{-z\bar{z}}$ in the measure,
as can be read from (\ref{dshwf}) and (\ref{dm}).
It is essentially a self-consistency
requirement, since the background charge is determined by the electron
distribution. On the other hand, for a transformation
$\epsilon(z)= \epsilon\,z$ with $\epsilon$ imaginary both terms in
(\ref{hatFz}) are equal instead of canceling one another.
This agrees with the fact that this transformation preserves the area,
since it is just a rotation.

We can in general split a conformal
transformation into two parts of different type in the following way.
Let us consider the unit circle and an arbitrary line
passing through the origin (the real axis, for example).
We define the two types of conformal transformations as the ones
that preserve either the circle or the line.%
\footnote{See the appendix for an algebraic treatment.}
Thus the former type of transformations consists of
the analytic continuation of diffeomorfisms of the circle $|z|=1$.
Similarly, the latter is the analytic continuation of diffeomorfisms
of the line. If we regard
the fluid droplet as confined within the unit circle, the first type is
such that preserves its boundary whereas the second type deforms it.

Since the dynamical degrees of freedom live on the edge of the sample,
it is natural to think of the diffeomorfisms of the circle as
the primary symmetry. Then one can analytically continue
a diffeomorfism of the circle to the interior (or exterior).
One can as well continue it as an area-preserving transformation.
Either way of continuing is achieved by considering $F$ or $\hat{F}$,
which coincide on the circle (as shown in the appendix).
On the contrary, the diffeomorfisms of the line have no physical
interpretation. For them $\hat{F}$ vanishes, as happened in
the particular case of a dilation. One must notice, however, that
this does not imply in general that the corresponding area-preserving
transformations vanish everywhere.

The essential factor that makes the difference between the symmetry
realizing as conformal or as area-preserving is e$^{-z{\bar z}}$.
This factor is not present in the ordinary representation, $\Psi(x,y)$.
For the conformal transformations that preserve the unit circle $D$
we have that
\be
\delta_{\epsilon}\int_D{\bar\psi}(\bar{z})\,\psi(z)\,d\mu = 0
\ee
If the density $\rho = |\Psi(x,y)|^2$ is constant over $D$
(as corresponds to an incompressible fluid), given that the size of
the liquid droplet is arbitrary, then
\be
\delta_{\hat{F}}\,(dx\,dy) = 0.
\ee
This is the condition for area-preserving diffeomorphisms.
Thus we see that unitarity and preservation of the area are closely related.
Of course, one cannot
prove with the present methods of first quantization
the existence of incompressibility, which is a property of
the ground state that involves interactions or, at least,
statistics. (A fluid of bosons would certainly not exhibit it.)
However, we know that it can be a simple property, essentially
entailed by the coupling of electrons to quantum fluxes embodied
by the Chern-Simons Lagrangian and patent in the mean-field theory
equation (\ref{mft}).

In conclusion,
$\psi(z)~\left({\bar\psi}(\bar{z})\right)$ transforms as a scalar
function under holomorphic (anti-holomorphic) diffeomorphisms. Thus,
the {\em first-quantized} Hall effect is a {\em classical} 2d CFT.
The holomorphic (anti-holomorphic) wave function can be expressed
in terms of another scalar field,
\be
\psi(z) = {\rm e}^{i\,\alpha\,\phi(z)}~~~\left( {\bar\psi}(\bar{z}) =
{\rm e}^{-i\,\alpha\,{\bar\phi}(\bar{z})} \right).
\label{vertex}
\ee
The field $\varphi(z,\bar{z})= \phi(z) + {\bar\phi}(\bar{z})$ satisfies
the free field equation
\be
\partial_z\partial_{\bar{z}}\,\varphi(z,\bar{z}) = 0.
\ee
Hence we make connection with the simplest 2d CFT, namely,
the Coulomb gas.
The classical Coulomb gas theory is insufficient to provide
an adequate description of the fractional QHE. In particular, it cannot
account for the existence of incompressible ground states at preferred
densities. Second quantization,
appropriate for the many interacting electron system, leads to
a fully quantum 2d CFT.
The realization of conformal symmetry
may have a central charge, which in the simplest Coulomb-gas theory
is $c=1$.
Excitations are described by vertex operators, corresponding to
(\ref{vertex}), and their statistical properties can change
according to the value of $\alpha$.
Some of these matters have already been discussed elsewhere;
see, for instance, ref. \cite{NNG}.

It is important to remark that unitarity of the realization of
conformal symmetry in the QHE is not related to the usual notion
of unitarity in conformal field theory, as was shown in section 4. It
is rather related to modular invariance, according to section 5.
It is thus possible to
consider non-unitary conformal models as candidate states for
the fractional QHE.

Finally, let us point out that there is another realization of 2d CFT in
the QHE. It originates in the idea of edge waves and
is formulated as a $1+1$ CFT living on the cylinder swept by
the boundary of the sample with time. Its connection to the euclidean
2d CFT at fixed time can be made with the methods of section 5.

\vspace{8mm}
\noindent
{\large \bf Acknowledgments}\\[2mm]
\nopagebreak
I am grateful to F. Alexander Bais, C{\'e}sar G{\'o}mez, J. Mateos Guilarte,
Germ{\'a}n Sierra, Michael Stone and Gillermo Zemba for conversations and
to Alfonso Ramallo for correspondence on Chern-Simons theory.

\appendix

\section{Appendix}

Let us formulate the splitting of a conformal transformation into
circle and line-preserving parts. The equations of the circle
and the line are $z{\bar z} =1$ and $z/{\bar z} =1$, respectively.
Under a conformal transformation
\bea
\delta z = \epsilon(z),  \\
\delta {\bar z} = \bar{\epsilon}({\bar z})
\eea
so
\bea
\delta (z\bar{z}) = \bar{z}\,\epsilon(z) + z\,\bar{\epsilon}({\bar z}),
\label{circle}  \\
\delta (z/\bar{z}) = \frac{\bar{z}\,\epsilon(z)-z\,\bar{\epsilon}({\bar z})}
                     {z^2}                 \label{line}
\eea
on the circle and line, respectively. From these formulas follow
the properties of $F$ or $\hat F$ mentioned in the text.

Introducing the Laurent expansion
of $\epsilon(z)$,
$$
\epsilon(z) = \sum_{-\infty}^{\infty} \epsilon_n\,z^{n+1},
$$
we can express that the variation (\ref{circle}) vanish as
\be
\epsilon_n + \bar{\epsilon}_{-n} = 0.     \label{cond}
\ee
In particular, if we want the transformation to be regular at
the origin, namely, $\epsilon_n = 0$ for $n<-1$, we obtain from
(\ref{cond}) that the only non-vanishing $\epsilon_n$ occur for
$n = -1,0,1$ and they further satisfy two equations. The total
number of real parameters is three and they are the well-known
projective transformations of the unit circle.

The condition (\ref{line}) just implies that
\be
\epsilon(z) = \bar{\epsilon}(z),
\ee
that is the transformation is real (all $\epsilon_n$ real).

\end{document}